\documentclass[%
 reprint,
 superscriptaddress,
 preprintnumbers,
 nofootinbib,
 amsmath,amssymb,
 aps
]{revtex4-2}

\usepackage[T1]{fontenc} 
\usepackage[utf8]{inputenc}
\usepackage[USenglish]{babel}
\usepackage{soul}
\usepackage{cancel}
\usepackage[normalem]{ulem}

\usepackage{multirow}
\usepackage[dvipsnames]{xcolor}
\usepackage{graphicx} 
\usepackage{mathtools}
\usepackage{float}
\usepackage{gensymb}

\usepackage{xcolor} 
\usepackage{amsmath,amsfonts,amssymb} 
\usepackage{siunitx}
\usepackage{tensor}
\usepackage{caption} 
\usepackage{subcaption}
\usepackage{dsfont}
\usepackage{braket}
\usepackage{dcolumn}
\usepackage{bm}
\DeclareMathAlphabet{\mathpzc}{OT1}{pzc}{m}{it}

\usepackage[bookmarks=true, pdfpagelabels=true, linkcolor=black, urlcolor=black, citecolor=blue, anchorcolor=blue]{hyperref}

\usepackage[capitalise,english]{cleveref}

\newcommand{\be}{\begin{equation}}
\newcommand{\ee}{\end{equation}}
\newcommand{\beq}{\begin{equation}}
\newcommand{\eeq}{\end{equation}}

\def\taup{ {\tau^+} }
\def\taum{{ \tau^-} }
\def\taupm{{ \tau^\pm} }
\def\ptaup{p_{\tau^+}^\mu}
\def\ptaum{p_{\tau^-}^\mu}

\def\pip{{\pi^+}}
\def\pim{{\pi^-}}
\def\pipm{{\pi^\pm}}
\def\ppip{p_{\pi^+}^\mu}
\def\ppim{p_{\pi^-}^\mu}

\def\ket{\rangle}
\def\bra{\langle}

\def\be{\begin{equation}}
\def\ee{\end{equation}}
\def\ba{\begin{eqnarray}}
\def\ea{\end{eqnarray}}

\def\Tr{\,{\rm Tr}\,}

\def\r{\rho}
\def\a{\alpha}

\def\g{\gamma}

\def\m{\mu}
\def\n{\nu}

\def\s{\sigma}

\makeatletter
\@ifundefined{textcolor}{}
{
 \definecolor{BLACK}{gray}{0}
 \definecolor{WHITE}{gray}{1}
 \definecolor{RED}{rgb}{1,0,0}
 \definecolor{GREEN}{rgb}{0,1,0}
 \definecolor{BLUE}{rgb}{0,0,1}
 \definecolor{CYAN}{cmyk}{1,0,0,0}
 \definecolor{MAGENTA}{cmyk}{0,1,0,0}
 \definecolor{YELLOW}{cmyk}{0,0,1,0}
}

\makeatother

\begin{document}

\title{Quantum information and CP measurement \\
in $H \to \tau^+ \tau^-$ 
at future lepton colliders}

\author{Mohammad Mahdi Altakach}
\email{altakach@lpsc.in2p3.fr}
\affiliation{%
Institute of Theoretical Physics, Faculty of Physics,
University of Warsaw, ul.~Pasteura 5, PL-02-093 Warsaw, Poland%
}
\affiliation{%
Laboratoire de Physique Subatomique et de Cosmologie,
Universit\'{e} Grenoble-Alpes, CNRS/IN2P3, Grenoble INP, 38000 Grenoble, France%
}

\author{Priyanka Lamba}
\email{priyanka.lamba@fuw.edu.pl}
\affiliation{%
Institute of Theoretical Physics, Faculty of Physics,
University of Warsaw, ul.~Pasteura 5, PL-02-093 Warsaw, Poland%
}

\author{Fabio Maltoni}
\email{fabio.maltoni@uclouvain.be}
\affiliation{Centre for Cosmology, Particle Physics and Phenomenology (CP3),  Universit\'e catholique de Louvain, 1348 Louvain-la-Neuve, Belgium}
\affiliation{Dipartimento di Fisica e Astronomia, Universit\`a di Bologna and INFN,
Sezione di Bologna, via Irnerio 46, 40126 Bologna, Italy}

\author{Kentarou Mawatari}
\email{mawatari@iwate-u.ac.jp}
\affiliation{%
Faculty of Education, Iwate University, Morioka, Iwate 020-8550, Japan%
}

\author{Kazuki Sakurai}
\email{kazuki.sakurai@fuw.edu.pl}
\affiliation{%
Institute of Theoretical Physics, Faculty of Physics,
University of Warsaw, ul.~Pasteura 5, PL-02-093 Warsaw, Poland%
}

\begin{abstract}
We introduce a methodology and investigate the feasibility  of measuring quantum properties of tau lepton pairs in the $H \to \tau^+ \tau^-$ decay
at future lepton colliders.
In particular, observation of entanglement, steerability and violation of Bell inequalities are examined for the ILC and FCC-ee.  
We find that detecting quantum correlation
crucially relies on precise reconstruction of the tau lepton rest frame and 
a simple kinematics reconstruction does not suffice due to the finite energy resolution of the colliding beams and detectors.
To correct for energy mismeasurements,
a log-likelihood method is developed that incorporates the information of impact parameters of tau lepton decays.
We demonstrate that an accurate measurement 
of quantum properties 
is possible with this method. As a by-product,
we show that a novel model-independent test of CP violation can be performed and the CP-phase of $H \tau \tau$ interaction can be constrained with an accuracy comparable to dedicated analyses, i.e., up to $ 7.9\degree$ and $5.4\degree$
at ILC and FCC-ee, respectively. 
\end{abstract}


\maketitle

\section{Introduction}

After almost one century since it was named by Born and an outstanding number of  predictions and experimental confirmations, quantum mechanics has become the founding aspect of all theories aiming to describe phenomena at the fundamental level. 
Together with special relativity, it provides the foundation of quantum field theory, on which the Standard Model (SM) of particle physics has been built.  
The most peculiar (and spectacular) trait of quantum mechanics is \emph{entanglement} \cite{epr,schrodinger_1935},  
a type of correlations between two (or more) subsystems that can survive even if they are spacelike separated. After having puzzled physicists for decades, entanglement has now become part of our everyday life, being the key to many advanced technologies, such as quantum computation, cryptography and teleportation (see e.g.\ \cite{nielsen00}). 
The nature of entanglement and quantum correlations has been studied intensively with the methods of quantum information theory, revealing  multiple levels of quantum correlations within entangled states.
In the strongest end, a correlation exists that  no classical system can account for it. 
Such a correlation can be detected as violation of Bell-type inequalities \cite{Bell:1964kc,Werner-Wolf-2001,Clauser:1969ny}.
In the low energy regime, 
violation of those inequalities 
has been observed in a number of experiments 
\cite{PhysRevLett.49.91, 
PhysRevLett.79.1,
Belle:2007ocp,
PhysRevLett.117.050402,
Steinhauer:2015saa,
Pfaff,
PhysRevA.94.012108}.\footnote{Relatively recently, ``loophole-free'' tests of Bell-type inequalities have been performed \cite{Hensen:2015ccp,
PhysRevLett.115.250401,
PhysRevLett.115.250402}
and the violation has been observed.}

High energy colliders, such as the Large Hadron Collider (LHC) at CERN, provide a unique and interesting environment to test  entanglement and other quantum correlations at the highest scales/shortest distances.  
Recently, tests of entanglement and Bell-type inequalities at the LHC have been proposed 
in the final states of $t \bar t$ \cite{Afik:2020onf,Fabbrichesi:2021npl,Severi:2021cnj,Afik:2022kwm,Aguilar-Saavedra:2022uye,Afik:2022dgh} and
a pair of weak bosons \cite{Barr:2021zcp,Barr:2022wyq,Aguilar-Saavedra:2022mpg,Aguilar-Saavedra:2022mpg,Aguilar-Saavedra:2022wam,Ashby-Pickering:2022umy}.
A theoretical discussion of entanglement in the production of a pair of photons and tau leptons has also been given \cite{Fabbrichesi:2022ovb}. 
The effect of beyond the Standard Model physics on entanglement measurements has been studied in the Standard Model Effective Field Theory (SMEFT) framework \cite{Severi:2022qjy,Aoude:2022imd}.

In this paper we introduce a new  methodology and study the feasibility to measure various quantum correlations within the tau  pairs in $H \to \tau^+ \tau^-$ at future lepton colliders, in particular, the International Linear Collider (ILC) \cite{Baer:2013cma,Behnke:2013lya}
and the Future Circular $e^+ e^-$ Collider (FCC-ee) \cite{FCC:2018evy}. 
Our aim is to access observables that signal different levels of quantum correlations: entanglement, steering and Bell-nonlocality, which we will define and discuss in detail in the next section.
We find that, employing standard reconstruction methods,  
an accurate measurement of quantum correlations is quite challenging even at lepton colliders
because of the presence of neutrinos in the final state
and finite beam and detector energy resolutions.
To improve the measurement accuracy, 
a log-likelihood method is proposed that incorporates the information of the impact parameters of tau lepton decays.
The spin correlation of tau lepton pairs in $H \to \tau^+ \tau^-$ is sensitive to the CP-phase of the $H \tau \tau$ interaction.
Exploiting this fact, we propose a model-independent test of CP violation.
We estimate the expected resolution of the CP-phase at the ILC and FCC-ee  
via quantum property measurements.

The paper is organised as follows.
In the next section we review the main properties of a bi-qubit system and introduce the notion of three types of quantum correlations: entanglement, steerability and Bell-nonlocality. 
We discuss their relations
and define some observables that are sensitivies to  each type of correlation.
In section \ref{sec:Htautau} we study the quantum state of the di-tau system in $H \to \tau^+ \tau^-$ with general $H \tau \tau$ interactions.  
Section \ref{sec:measurement}
describes our strategy to measure the quantum spin correlation of the tau pairs in $H \to \tau^+ \tau^-$.
Our assumptions on the future $e^+ e^-$ colliders are spelled out in Section \ref{sec:leptoncolliders}.
In section \ref{sec:analysis}
we describe the details of our Monte Carlo (MC) simulation and event analysis.  
The result of the quantum property measurements (based on MC simulations) is also shown. As a by-product of our analysis, 
we propose a novel model-independent test of CP violation 
and the resolution for the CP-phase of the
$H \tau \tau$ coupling is estimated in section \ref{sec:CP}.
Section \ref{sec:Conclusions} is devoted to the conclusion and discussion.

\section{Quantum nature of biparticle systems}
\label{sec:QM}

\subsection{Entanglement}
\label{sec:entangle}

The Hilbert space of the spin 1/2 biparticle system is spanned by the four basis kets, 
$\{ |1 \ket, |2 \ket, |3 \ket, |4 \ket \} 
= \{ | +,+ \ket, | +,- \ket, | -,+ \ket, | -,- \ket\} $. On the RHS, $| m_A, m_B \ket$ is a simultaneous eigenstate of the spin $z$-component, $\hat s_z^I$, of particles $A$ and $B$, respectively, i.e.\ $\hat s^A_z | \pm, m_B \ket = \pm | \pm, m_B \ket$
and 
$\hat s^B_z | m_A, \pm \ket = \pm  | m_A, \pm \ket$.\footnote{We use the spin operators, $\hat s_i^I$, that are scaled by a factor $2/\hbar$ compared to the usual ones, so that the eigenvalues are $\pm 1$.}
In this basis the density operator, $\rho$, for a general mixed state is represented by a $4 \times 4$ matrix 
\be
\rho = \frac{1}{4} \left[ 
{\mathbf 1} \otimes {\mathbf 1} 
+ B_i \sigma_i \otimes {\mathbf 1}  
+ \overline B_i {\mathbf 1}  \otimes \sigma_i 
+ C_{ij} \sigma_i \otimes \sigma_j \right]\,,
\label{eq:rho_expand}
\ee
where the summation of $i,j = 1,2,3$ indices is implicit and $\s_i$ are the Pauli matrices.
The physical density matrix is Hermitian, $\Tr(\rho) = 1$ and positive definite.  
The Hermiticity condition implies the coefficients $B_i$, $\bar B_i$ and $C_{ij}$ are real.
The expectation value of a physical observable $\hat O$ is 
given as $\bra \hat O \ket = \Tr( \hat O \rho)$. It follows 
$B_i = \bra \hat s_i^A \ket$, $\overline B_i = \bra \hat s_i^B \ket$
and 
$C_{ij} = \bra \hat s_i^A \hat s_j^B \ket$, giving clear interpretation to these coefficients:
$B_i$ ($\bar B_i$) is the spin polarization of $A$ ($B$) and $C_{ij}$ is the spin correlation.
From this interpretation, it is apparent that the magnitude of these coefficients is less than or equal to 1.

If a density matrix can be written in the form 
\be
\rho = \sum_k p_k \, \rho^A_k \otimes \rho^B_k\,,
\ee
with $p_k \ge 0$ and $\sum_k p_k = 1$,
the state is said to be {\it separable}.
Conversely, non-separable states are called {\it entangled}.
A sufficient condition of entanglement
is obtained by taking a partial transpose for 
the $B$ part
\be
\rho^{T_B} \equiv \sum_k p_k \, \rho^A_k \otimes 
\left( \rho^B_k \right)^T\,.
\ee
If the state is separable, $\rho^{T_B}$ must still be
a physical density matrix, in particular positive definite. 
If one finds a negative eigenvalue for $\rho^{T_B}$, the system must therefore be entangled.  
This condition, known as the Peres-Horodecki criterion \cite{Peres:1996dw,Horodecki:1997vt},
is also a necessary condition of entanglement for spin $1/2$ biparticle systems.
A simple sufficient condition that  $\rho^{T_B}$
has a negative eigenvalue is given by (see e.g.\ \cite{Afik:2020onf, Aguilar-Saavedra:2022uye})
\be
E \equiv \max_i \left\{ \left| \Tr( C ) - C_{ii}  \right| - C_{ii} \right\} > 1\,.
\label{eq:E}
\ee

A more quantitative measure of the entanglement of two qubits is given by the \emph{concurrence} \cite{Wootters:1997id}, which is defined by
\be
{\cal C}[\rho] \equiv {\rm max}(0, \eta_1 - \eta_2 - \eta_3 - \eta_4),
\label{eq:concurrence}
\ee
where $\eta_i$ is the eigenvalues of a matrix ${\cal R} \equiv \sqrt{ \sqrt{\rho} \tilde \rho \sqrt{\rho}}$
in the descendent order, $\eta_i \geq \eta_j$ ($i < j$), and $\tilde \rho \equiv (\sigma_2 \otimes \sigma_2) \rho^* (\sigma_2 \otimes \sigma_2)$.
Although ${\cal R}$ is generally non-Hermitian, 
the eigenvalues are real and non-negative.
The concurrence takes a value in the range 
$0 \leq {\cal C}[\rho] \leq 1$.
For separable states ${\cal C}[\rho] = 0$,
while ${\cal C}[\rho] = 1$ for maximally entangled states.

\subsection{CHSH inequality}
\label{sec:CHSH}

Consider the following experiment. 
Four identical (statistical) samples of spin $1/2$ biparticle systems are prepared, labeled by $\rho_{ab}$, $\rho_{a'b}$,
$\rho_{ab'}$ and $\rho_{a'b'}$
($\rho_{ab} = \rho_{a'b} = \rho_{ab'} = \rho_{a'b'} \equiv \rho$).
Alice and Bob measure the spin components of particles $A$ and $B$, respectively. 
Their measurements are spacelike separated.
Alice can measure the spin in the directions of either $\bf a$ or $\bf a'$ 
while Bob can choose between $\bf b$ or $\bf b'$.
For the samples labeled by $\rho_{aX_b}$ ($X_b = b, b'$), Alice uses $\bf a$ direction for her measurements, while she uses $\bf a'$ direction
for the samples labeled by $\rho_{a'X_b}$.
Similarly, Bob measures the spin in $\bf b$ ($\bf b'$) direction for the samples labeled by $\rho_{X_a b(b')}$ ($X_a = a, a'$).

Let ${\cal M}_A$ and ${\cal M}_B$ be 
all possible measurement axes of Alice and Bob, respectively.
If their measurement axes are $\cal{A} \in {\cal M}_A$
and $\cal{B} \in {\cal M}_B$,
the conditional probability
of observing the outcomes $a = \pm 1$ and $b = \pm 1$, respectively, 
is written by
$p(a,b|  \cal{A}, \cal{B} )$.
If this probability can be written in terms of 
a set of ``hidden'' variables $\lambda$ with
probability distribution $P(\lambda)$ as
\be
p(a,b| {\cal A}, {\cal B} ) \,=\,
\sum_{\lambda} \, P( \lambda ) \, p(a | {\cal A}, \lambda) \, p(b| {\cal B}, \lambda)\,,
\ee
for all $\cal{A}$, $\cal{B}$, $a$, $b$,
the state $\rho$ is said to be {\it Bell}-{\it local} \cite{Bell:1964kc, Werner-Wolf-2001}, where $p(a | {\cal A}, \lambda)$ and $p(b| {\cal B}, \lambda)$ are the
conditional probabilities for Alice and Bob, respectively, when hidden variables take value $\lambda$.

After the measurements,  various spin correlations can be computed.  In particular, we are interested in the following quantity \cite{Clauser:1969ny}: 
\be
R_{\rm CHSH} = \frac{1}{2} \left| 
\bra \hat s^A_a \hat s^B_b \ket
- \bra \hat s^A_a \hat s^B_{b'} \ket
+ \bra \hat s^A_{a'} \hat s^B_{b} \ket
+ \bra \hat s^A_{a'} \hat s^B_{b'} \ket
\right|\,.
\label{eq:R}
\ee
It has been shown 
that for any Bell-local state
$R_{\rm CHSH}$ is bounded from above by 1
for any four measurement axes.
Namely,
\be
R^{\rm max}_{\rm CHSH} \equiv \max_{a,a',b,b'} R_{\rm CHSH} \leq 1~~~({\rm Bell\text{-}local})\,,
\label{CHSH}
\ee
where the maximum is taken over all unit vectors, $\bf a$, $\bf a'$, $\bf b$ and $\bf b'$.
This bound is known as the Clauser-Horne-Shimony-Holt (CHSH) inequality \cite{Clauser:1969ny},
which is one of the Bell-type inequalities \cite{Bell:1964kc}.
Experimental observation of violation the CHSH inequality would confirm Bell-nonlocality
and falsify all local-real hidden variable theories.
Quantum mechanics, on the other hand, can violate the CHSH inequality up to $\sqrt{2}$:
\be
R^{\rm max}_{\rm CHSH} \leq \sqrt{2}~~~({\rm QM}) \,.
\ee
This quantum mechanical bound is
known as the Tsirelson bound \cite{Tsirelson}.

For two-qubit systems, 
$R^{\rm max}_{\rm CHSH}$
can be analytically calculated as
\be
R^{\rm max}_{\rm CHSH} = \sqrt{\mu_1 + \mu_2} \,,
\label{eq:Sformula}
\ee
where $\mu_i$ ($\mu_1 \geq \mu_2 \geq \mu_3$) are the eigenvalues of the matrix $C^T C$. 
The set of unit vectors, $\bf a_*$, $\bf a'_*$, $\bf b_*$ and ${\bf b'_*}$, which maximises $R_{\rm CHSH}$ is given by
\ba
&& {\bf a_*} = \frac{1}{\sqrt{\mu_1}} C \, {\bf d}_1,~~~~
{\bf a'_*} = \frac{1}{\sqrt{\mu_2}} C \, {\bf d}_2,
\nonumber \\
&& {\bf b_*} \,=~~ \cos \varphi \, {\bf d}_1 + \sin \varphi \, {\bf d}_2, 
\nonumber \\
&& {\bf b'_*} \,= - \cos \varphi \, {\bf d}_1 + \sin \varphi \, {\bf d}_2, 
\label{eq:ab_max}
\ea
where $\varphi \equiv \arctan (\sqrt{\mu_1 / \mu_2})$
and
${\bf d}_1$, ${\bf d}_2$ are the normalised eigenvectors of
the matrix $C^T C$; $(C^T C) {\bf d}_i = \mu_i {\bf d}_i$.

\subsection{Steerability}
\label{sec:steer}

Non-locality of quantum states was first pointed out 
by Einstein-Podolsky-Rosen (EPR) in 1935 \cite{Einstein:1935rr}.
Schr{\" o}dinger reacted to this work
and introduced a concept called \emph{steering} together with entanglement in his 1935 paper \cite{schrodinger_1935}.
In the previous biparticle system of Alice and Bob,
steering by Alice is Alice's ability to affect 
Bob's state by her measurement.
Although the concept is old, the formal definition of steering was found relatively recently \cite{PhysRevLett.98.140402,Jones2007}. 
The state $\rho$ is said to be \emph{steerable} by Alice 
if it is \emph{not} possible to write the probability distribution of measurement outcomes as
\be
p(a,b| {\cal A}, {\cal B} ) \,=\,
\sum_{\lambda} \, P( \lambda ) \, p(a | {\cal A}, \lambda) \, p_Q(b| {\cal B}, \lambda)\,,
\label{eq:steer1}
\ee
with
\be
p_Q(b| {\cal B}, \lambda) \equiv {\rm Tr}[ \rho_B(\lambda) F^{\cal B}_b],
\label{eq:steer2}
\ee
for all $\cal{A}$, $\cal{B}$, $a$ and $b$,
where $\rho_B(\lambda)$ is Bob's local state 
and $F_b^{\cal B}$ is Bob's positive operator valued measure \cite{nielsen00}.
An operational definition of steerability 
is also given in Appendix \ref{ap:steering}.

By definition, all Bell-nonlocal states are steerable.
Also, if states are separable, the probability of measurement outcomes can be written in the form of Eqs.~\eqref{eq:steer1} and \eqref{eq:steer2}.
Namely, the following hierarchy is established \cite{PhysRevLett.98.140402}:
\be
{\rm Entangled}
~\supset~
{\rm Steerable}
~\supset~
{\rm Bell}{\text -}{\rm nonlocal}\,.
\ee

\section{Quantum and CP properties of $H \to \tau^+ \tau^-$}
\label{sec:Htautau}

In the $H \to \tau^+ \tau^-$ decay, the spins of two tau leptons form a two-qubit system and can be used to test various quantum information properties. 
We now calculate the observables introduced in the previous section for the two-qubit system in  the $H \to \tau^+ \tau^-$ decay.

A generic interaction between a Higgs boson and tau leptons can be written as
\be
{\cal L} \,\ni\, - \frac{m_{\tau}}{v_{\rm SM}} \kappa
\, H \bar \psi_\tau (\cos \delta + i \g_5 \sin \delta) \psi_\tau \,,
\label{eq:Htautau}
\ee
where $m_{\tau}$ and $v_{\rm SM}$ are the tau lepton mass and the SM Higgs vacuum expectation value, respectively.
The real parameters $\kappa \in {\mathbb R}_+$ and $\delta \in [0, 2 \pi]$ describe the magnitude of the Yukawa interaction and the CP phase.
Within this parametrization, the Standard Model corresponds to  $(\kappa, \delta) = (1,0)$.

The spin density matrix for the two tau leptons 
is given by
\be
\rho_{m n, \bar m \bar n} = \frac{ {\cal M}^{* n \bar n} {\cal M}^{m \bar m} }{ \sum_{m \bar m}  \left| {\cal M}^{ m \bar m} \right|^2 },
\ee
where
\be
{\cal M}^{m \bar m} = c \, \bar u^m(p) (\cos \delta + i \g_5 \sin \delta) v^{\bar m}(\bar p)
\ee
is the amplitude of $H \to \tau^+ \tau^-$
and $c = -i \kappa m_\tau/v_{\rm SM}$. 
Here, $p^\mu = (\frac{m_{H}}{2}, 0,0,p_z)$ and 
$\bar p^\mu = (\frac{m_{H}}{2}, 0,0, -p_z)$ are the momenta of $\tau^-$
and $\tau^+$, respectively, in the Higgs boson rest frame.
The indices $m,n$ ($\bar m, \bar n$) label the $\tau^{-(+)}$ spin in the direction of the $z$-axis (the direction of $\tau^-$ momentum).
A straightforward calculation leads to \cite{Bernreuther:1997gs,Desch:2003rw}
\be
\rho_{m n, \bar m \bar n} = 
\frac{1}{2}
\begin{pmatrix}
0 & 0 & 0 & 0 \\
0 & 1 & e^{-i2\delta} & 0 \\
0 & e^{i2\delta} & 1 & 0 \\
0 & 0 & 0 & 0 
\end{pmatrix}
\label{eq:rho_tau}
\ee
up to the term of the order of $m^2_{\tau}/m^2_H$.
On the RHS, the column ($m n$) and row ($\bar m \bar n$) are ordered as $(+,+), (+,-), (-,+), (-,-)$.
From this the expansion coefficients in Eq.\ \eqref{eq:rho_expand} can readily be obtained as
$B_i = \bar B_i = 0$ and
\be
C_{ij} = \begin{pmatrix}
\cos 2 \delta & \sin 2 \delta & 0  \\
-\sin 2 \delta & \cos 2 \delta  & 0  \\
0 & 0  & -1  \\
\end{pmatrix}\,.
\label{eq:Cmat}
\ee

The signature of entanglement \eqref{eq:E} is calculated to be  
\be
E(\delta) = 2 | \cos 2 \delta | + 1\,.
\ee
This is greater than 1 unless $\delta = \frac{\pi}{4}, \frac{3\pi}{4}, \frac{5\pi}{4}$ 
and reaches the maximum ($E(\delta)=3$) at $\delta = 0$ (SM), $\pi/2$ (CP-odd)
and $\pi$ (negative Yukawa coupling).

The concurrence is also calculable. 
Eq.\ \eqref{eq:rho_tau} leads to $\tilde \rho = \rho$ and ${\cal R} = \rho$.
It also implies $\eta_i = (1,0,0,0)$
and we therefore have ${\cal C}[\rho] = 1$. 
The $\tau^+ \tau^-$ pair is maximally entangled
regardless of the CP phase $\delta$ \cite{Fabbrichesi:2022ovb}.

For states with vanishing Bloch vectors, $B_i = \bar B_i = 0$, a convenient sufficient and necessary condition for steerability is known \cite{Jevtic_2015,Nguyen_2016,Xiao-Gang}.
The state is steerable if and only if $S[\rho] > 1$ 
with
\be
{\cal S}[\rho] \equiv \frac{1}{2 \pi} \int d \Omega_{\bf n} \sqrt{ {\bf n}^T C^T C {\bf n} }\,,
\label{eq:S}
\ee
where ${\bf n}$ is a unit vector to be integrated out.
In $H \to \tau^+ \tau^-$ we obtain ${\cal S}[\rho] = 2$ (steerable) from Eq.\ \eqref{eq:Cmat}.
We use ${\cal S}[\rho]$ as a measure of steering in the following sections.

The variable $R_{\rm CHSH}^{\max}$ can be calculated immediately  from Eq.\ \eqref{eq:Sformula} as
\be
R_{\rm CHSH}^{\rm max} = \sqrt{2},
\ee
which violates the classical bound and saturates the quantum mechanical one.
Since $R_{\rm CHSH}^{\rm max}$ is independent of $\delta$,
a test of Bell-nonlocality can be done regardless of the CP property of the $H \tau \tau$ interaction.

The state in Eq.\ \eqref{eq:rho_tau} is pure, i.e.\ $\Tr \rho^2 = 1$. 
The corresponding pure state can be found as \cite{Rouge:2005iy}
\be
| \Psi_{H \to \tau \tau}(\delta) \ket = \frac{1}{\sqrt{2}} \left( | +,- \ket + e^{i 2 \delta} | -,+ \ket \right)\,.
\label{eq:pure}
\ee
In the Standard Model ($\delta = 0$),
this state is the triplet state $(s,m) = (1,0)$,
where $s$ and $m$ are the magnitude and the $z$-component of the total spin, respectively.
This can be understood as follows.
Since the SM Higgs is CP-even scalar, the final state must have even parity and zero total angular momentum, $J^P = 0^+$,
provided the parity is conserved in the $H \tau \tau$ interaction. 
In the final state, the total parity is given by $P = (\eta_{\tau^-} \, \eta_{\tau^+}) \cdot (-1)^\ell$, where $\eta_{\tau^{-(+)}}$ is the intrinsic parity of $\tau^{-(+)}$ and $\ell$ is the orbital angular momentum.
The intrinsic parities of a fermion and its anti-fermion are opposite,
$(\eta_{\tau^-} \, \eta_{\tau^+}) = -1$, and the spin state of the final state must be $s = 0$ or 1.  
The only consistent choice to obtain $J^P = 0^+$ 
is $\ell = 1$ and $s=1$. 
The same line of argument leads to a conclusion that 
if $\tau^+ \tau^-$ are produced from the decay of a particle with $J^P = 0^-$
($\delta = \frac{\pi}{2}$),
the final state must have $\ell = s = 0$, 
namely, it must be the singlet state $\frac{1}{\sqrt{2}} (| +,- \ket - | -,+ \ket)$.
This observation is consistent with Eq.\ \eqref{eq:pure}.

\section{Measurement strategy}
\label{sec:measurement}

\begin{figure}[t!]
    \centering
	\includegraphics[width=0.60\linewidth]{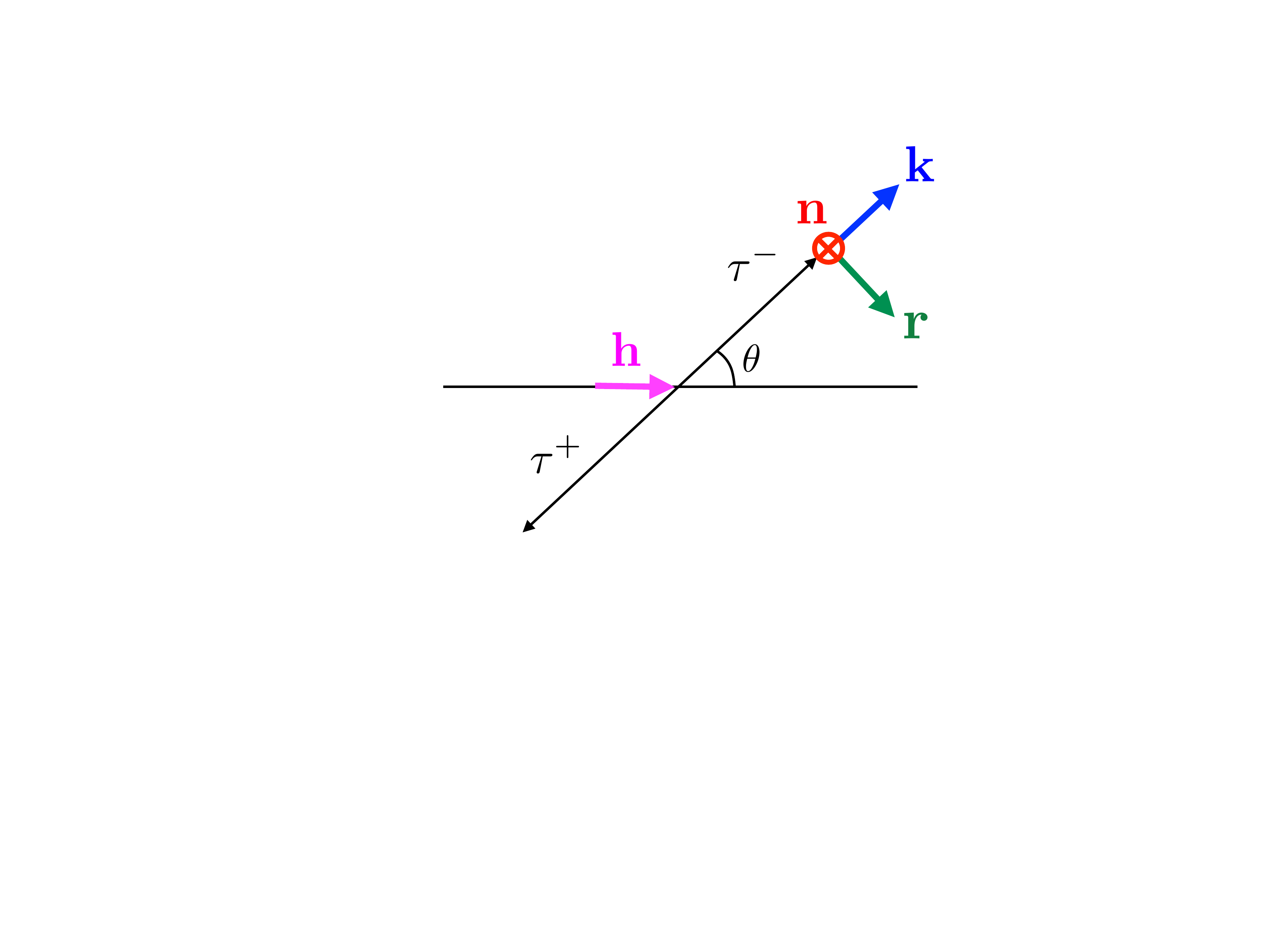}
	\caption{Helicity basis.
	}
	\label{fig:basis}
\end{figure}

The spin of  the tau leptons is not directly measureable at colliders.
What can be measured instead is the direction of a decay product with respect to the motion of the tau.
In order to sensibly compare the directions of decay products among different events, we adopt a coordinate system so-called the helicity basis \cite{Baumgart:2012ay}.
The three normalised basis vectors ($\bf r$, $\bf n$, $\bf k$) are introduced 
(see Fig.~\ref{fig:basis})
at the centre of mass frame of $\tau^+ \tau^-$
in the following way:
${\bf k}$ is the direction of $\tau^-$,
${\bf r}$ is on the plane spanned by ${\bf k}$ and ${\bf h}$,
which is the motion of the Higgs in the $\tau^+ \tau^-$ rest frame, and defined as ${\bf r} \equiv ({\bf h} - {\bf k} \cos \theta)/\sin \theta$ with $\cos \theta \equiv {\bf k} \cdot {\bf h}$,
and ${\bf n} \equiv {\bf k} \times {\bf r}$.

Suppose that at the rest frame of $\tau^{-}$ the tau spin is polarised into $\bf s$ direction ($|{\bf s}| = 1$).  
The $\tau^{-}$ decays into a decay mode, $f$, producing a detectable particle $d$.  
The conditional probability that the particle $d$ takes the direction $\bf u$ ($|{\bf u}| = 1$) when the $\tau^-$ spin is polarised in ${\bf s}$ direction is given by \cite{Bullock:1992yt}
\be
P({\bf u}|{\bf s}) \, = \, 1 + \alpha_{f,d} \,   {\bf s} \cdot {\bf u}\,,
\label{eq:prob}
\ee
with the normalisation $\int \frac{d \Omega}{4 \pi} P({\bf u}|{\bf s}) = 1$,
where $\alpha_{f,d} \in [-1, 1]$ is called the spin analyzing power.
For the CP counterpart, $(f,d) \xleftrightarrow{\rm CP} (\bar f, \bar d)$, $\alpha_{\bar f, \bar d} = - \alpha_{f, d}$.

We denote the $\tau^+$ polarization by $\bf \bar s$ ($|{\bf \bar s}| = 1$). 
The direction of its decay product, $d'$, measured at the rest frame of the $\tau^+$, is represented by a unit vector ${\bf \bar u}$.
We want to relate the spin correlation $\bra {\bf s} \otimes {\bf \bar s} \ket$
with the angular correlation $\bra {\bf u} \otimes {\bf \bar u} \ket$ since the latter is measureable. 
Using the probability distribution \eqref{eq:prob}, it is not hard to show (see Appendix \ref{ap:correlation})
\be
\bra u_a \bar u_b \ket \,=\, \frac{\alpha_{f,d} \alpha_{f',d'}}{9} \bra s_a \bar s
_b \ket  \,,
\label{eq:angle-to-spin}
\ee
where $u_a \equiv {\bf u} \cdot {\bf a}$, $\bar s_b \equiv {\bf \bar s} \cdot {\bf b}$, etc.\ are the components with respect to arbitrary unit vectors ${\bf a}$ and ${\bf b}$.
Using this relation, we can obtain $R_{\rm CHSH}$ in terms of the angular correlations:
\ba
&&R_{\rm CHSH} \,=\,  \frac{9}{2|\alpha_{f,d} \alpha_{f',d'}|} \times 
\nonumber \\
&&~~~\left| 
\bra u_a \bar u_b \ket
- \bra u_a \bar u_{b'} \ket
+ \bra u_{a'} \bar u_{b} \ket
+ \bra u_{a'} \bar u_{b'} \ket
\right|\,.~~~~~~
\label{eq:SCHSH_u}
\ea
In $H \to \tau \tau$, 
a set of four unit vectors that maximises $R_{\rm CHSH}$ can be chosen as
(see Eqs.~\eqref{eq:ab_max} and \eqref{eq:Cmat})
\ba
&&~~~~~~~~~~~~{\bf a}_* = {\bf r},~~~~{\bf a'}_* = {\bf n}, 
\nonumber \\
&&{\bf b}_* = \frac{1}{\sqrt{2}} \left( {\bf n} + {\bf r} \right),~~
{\bf b'}_* = \frac{1}{\sqrt{2}} \left( {\bf n} - {\bf r} \right)\,.
\label{eq:max_abab}
\ea
We use the above unit vectors and 
consider a direct measurement of $R^*_{\rm CHSH} \equiv R_{\rm CHSH}( {\bf a}_*, {\bf a'}_*, {\bf b}_*, {\bf b'}_*)$ to test the Bell-nonlocality in section \ref{sec:analysis}.

From Eq.\ \eqref{eq:prob}, one can also show \cite{Bernreuther:2004jv}
\be
\frac{1}{\sigma}\frac{d \sigma}{d(u_a \bar u_b)} \,=\, 
\frac{1 + \alpha_{f,d} \alpha_{f',d'} C_{ab} \, u_a \bar u_b}{2}
\ln \left( \frac{1}{ u_a \bar u_b } \right) \,.
\ee
This allows us to measure the $C_{ab}$ component by fitting the $\frac{d \sigma}{d(u_a \bar u_b)}$ distribution 
with the function on the RHS \cite{Severi:2021cnj}.
It has been pointed out that
the components of the $C$-matrix can also be measured from the forward-backward asymmetry 
\cite{Aguilar-Saavedra:2022uye}
\be
C_{ab} = \frac{4}{-\alpha_{f,d} \alpha_{f',d'}} 
\frac{N( u_a \bar u_b > 0 ) - N( u_a \bar u_b < 0 )}{N( u_a \bar u_b > 0 ) + N( u_a \bar u_b < 0 )} \,.
\label{eq:C_N}
\ee
The simplest approach to measure the $C$-matrix is to use Eq.~\eqref{eq:angle-to-spin}:
\be
C_{ab} = \bra s_a \bar s_b \ket
= \frac{9}{\alpha_{f,d} \alpha_{f',d'}}
\bra u_a \bar u_b \ket\,.
\label{eq:C_from_uu}
\ee
We have tested the above three approaches to measure $C_{ab}$ and found very similar results.  
Our final result in the following sections is based on the simplest method \eqref{eq:C_from_uu} since it has given
the most precise result among the three approaches.

For the steering measurement, we calculate ${\cal S}[\rho]$ by directly performing the integral in Eq.\ \eqref{eq:S} with the measured $C$-matrix. 

In the Standard Model ($\delta = 0$), the $C$-matrix in the helicity basis is given by
\be
C_{rr} = C_{nn} = 1,~~C_{kk} = -1,~~C_{ij}=0~(i \neq j) 
\ee
and the entanglement signature becomes  
\be
E = E_k \equiv C_{rr} + C_{nn} - C_{kk}\,.  
\ee

There is a way to measure this combination directly \cite{Aguilar-Saavedra:2022uye}.
We introduce a metric $\eta_k = {\rm diag}(1,1,-1)$
and define $\cos \theta_k \equiv {\bf u}^T \eta_k {\bf \bar u} = u_r \bar u_r + u_n \bar u_n - u_k \bar u_k$.
This quantity distributes as 
\be
\frac{1}{\sigma} \frac{d \sigma}{d \cos \theta_k}
= \frac{1}{2} \, \big( \, 1 - \alpha_{f,d} \alpha_{f',d'} E_k \cos \theta_k \, \big) 
\ee
and $E_k$ can be measured as a forward-backward asymmetry 
\be
E_k = \frac{6}{-\alpha_{f,d} \alpha_{f',d'}} 
\frac{N( \cos \theta_k > 0 ) - N( \cos \theta_k < 0 )}{N( \cos \theta_k > 0 ) + N( \cos \theta_k < 0 )} \,.
\label{eq:Ek_N}
\ee
In our numerical analysis, we calculate $E_k$ with 
Eq.\ \eqref{eq:Ek_N}.
Entanglement is detected if $E_k > 1$.

For the concurrence measurement, we assume $R = \rho$,
as suggested in Eq.\ \eqref{eq:rho_tau}.
Using $\Tr \rho = 1$, the concurrence can be expressed by ${\cal C}[\rho] = {\max}(0, 2 \eta_1 - 1)$,
where $\eta_1$ is the largest eigenvalue of $\rho$.
Assuming all off-diagonal entries of $C$, 
except for $C_{rn}$ and $C_{nr}$, vanish,
we have
\be
{\cal C}[\rho] = {\rm max} \left[ 0,\, \frac{D_{+} + C_{kk} - 1}{2},\,
\frac{D_{-} - C_{kk} - 1}{2} 
\right]
\label{eq:crho_tau}
\ee
with $D_{\pm} \equiv \sqrt{(C_{rn} \pm C_{nr})^2 + (C_{rr} \mp C_{nn})^2}$.
In our measurement, we construct the concurrence 
from measured $C$-matrix entries 
using the above expression. 
${\cal C}[\rho] > 0$ signals a formation of entanglement and
${\cal C}[\rho] = 1$ implies maximally entangled states.

\section{High energy $e^+ e^-$ colliders}
\label{sec:leptoncolliders}

For testing entanglement and Bell-nonlocality in $H \to \tau^+ \tau^-$,
high energy $e^+ e^-$ colliders have two main advantages over hadron colliders.
First, background is much smaller for lepton colliders.
At $pp$ colliders, the main production mode is $g g \to H \to \tau^+ \tau^-$, which is loop-induced.
This final state is generally contaminated by the tree-level $q \bar q \to Z^* \to \tau^+ \tau^-$ process, in which the final state tau pair belongs to a different quantum state than in the signal.
The main handle for signal/background separation is the invariant mass of the visible decay products of two tau leptons, $m_{\rm vis}(\tau^+ \tau^-)$.  
However, due to presence of neutrinos in tau decays, the $m_{\rm vis}(\tau^+ \tau^-)$ distributions have long tails and the signal and background distributions overlap significantly.
A usual practice to overcome this problem is to try to reconstruct the tau momenta by making some assumption on the neutrino momenta, based on either kinematics (e.g.~collinear approximation) or the knowledge of the Standard Model (e.g.~likelihood approach). 
However, this is not an option here, since our aim is to measure the angular distribution. 
Assuming the Standard Model distribution simply defeats the purpose of the measurement.  

At $e^+ e^-$ colliders, the main production channel near the threshold, $\sqrt{s} \sim (m_H + m_Z)$, is
$e^+ e^- \to Z H$ followed by $Z \to q \bar q/ \ell^+ \ell^-$ and $H \to \tau^+ \tau^-$.
The main background is $e^+ e^- \to Z \tau^+ \tau^-$,
where the pair of taus comes from an exchange of $\gamma^*/Z^*$.
Unlike hadron colliders, the full 4-momentum, $P^\mu_{\rm in}$, of the initial state ($e^+ e^-$ pair)
is precisely known at lepton colliders.  
From this and the measured $Z$-boson momentum, $p_Z^\mu = (p_{q/\ell^-} +\, p_{\bar q/\ell^+})^\mu$,
one can reconstruct the Higgs momentum as
\be
p_H^\mu = P_{\rm in}^\mu - p_Z^\mu
\ee
in a good accuracy independently from the Higgs decays.
The distribution of the recoil mass, $m_{\rm recoil} = \sqrt{(P_{\rm in} - p_Z)^2}$,
therefore sharply peaks at the Higgs mass in the signal \cite{Yan:2016xyx}.
By selecting events that fall within an narrow window, $|m_{\rm recoil} - m_{H}| < 5$ GeV, one can achieve background/signal $\sim 0.05$ with a signal efficiency of $93\%$ and $96\%$ for the ILC and FCC-ee, respectively.  

The second advantage of $e^+ e^-$ colliders over hadron colliders is the ability of reconstructing two tau momenta by solving kinematical constraints.
This is possible thanks to the fact that 
the initial state 4-momentum, $P^\mu_{\rm in}$, is known in a good precision.
This is important for the $C$-matrix measurement  
and the Bell inequality test since they are 
based on the angular distributions of ${\bf u}$ and ${\bf \bar u}$, which must be performed at the rest frames of $\tau^-$ and $\tau^+$, respectively.  
Since taus are heavily boosted, a small error on the tau momentum leads to a large error on the angular distribution when boosted to the tau rest frame.
Precise reconstruction of the tau momenta is therefore crucial for the $C$-matrix measurement and Bell inequality test.


We consider two benchmark collider scenarios 
labelled by ``ILC'' \cite{Baer:2013cma} and ``FCC-ee'' \cite{FCC:2018evy}.
The relevant parameters we use in our simulation
are listed in Table \ref{tb:colliders}.
\begin{table}[h!]
\begin{center}
\begin{tabular}{r | c | c} 
  & ~~~ILC~~~ & ~~~FCC-ee~~~ \\ [0.5ex] 
 \hline 
 energy (GeV) & 250 & 240  \\ [0.5ex]
 luminosity (ab$^{-1}$) & 3 & 5  \\[0.5ex]
 beam resolution $e^+$ (\%) & 0.18 & $0.83 \times 10^{-4}$  \\[0.5ex]
beam resolution $e^-$ (\%) & 0.27  & $0.83 \times 10^{-4}$  
\\[0.5ex]
$\sigma(e^+ e^- \to HZ)$ (fb) & 240.1  & 240.3 
\\[0.5ex]
\# of signal ($\sigma \cdot {\rm BR} \cdot L \cdot \epsilon$) & 385 & 663
\\[0.5ex]
\# of background ($\sigma \cdot {\rm BR} \cdot L \cdot \epsilon$) & 20 & 36
\end{tabular}
\caption{Parameters for benchmark lepton colliders 
\cite{Baer:2013cma, FCC:2018evy}.
Only the main background, $e^+ e^- \to Z \tau^+ \tau^-$,
is considered, where $\tau^+ \tau^-$ are produced from off-shell $Z/\gamma$.
The numbers of signal and background reported here include
the decay branching ratios and the efficiency of the event selection, $|m_{\rm recoil} - m_H| < 5$ GeV.
\label{tb:colliders}
}
\end{center}
\end{table}

\noindent
We notice that the beam energy resolution is significantly better for FCC-ee, which will have a significant impact on the Bell inequality test as we will see in the next section.
We assume the $e^+ e^-$ beams are unpolarised for both
ILC and FCC-ee.

\section{Event analysis and results}
\label{sec:analysis}

In our analysis, we focus on the tau decay modes:
\be
\tau^- \to \nu_\tau \pi^-,~~~~\tau^+ \to \bar \nu_\tau \pi^+
\label{eq:decay}
\ee
with ${\rm Br}(\tau^- \to \nu \pi^-) = 0.109$ \cite{Workman:2022ynf}.
For these decay modes, the spin analyzing power is maximum, $\alpha_{f,d} \alpha_{\bar f, \bar d} = -1$.
We generate signal and background events with {\tt MadGraph5\_aMC@NLO}
\cite{Alwall:2014hca} at leading order in the Standard Model, i.e.~$(\kappa, \delta) = (1,0)$.
We employ the {\tt TauDecay} package for $\tau$ decays~\cite{Hagiwara:2012vz}.
The beam energies are smeared according to the parameters in Table \ref{tb:colliders}.
All ``neutrinoless'' $Z$-boson decay modes, $Z \to x \bar x$ with $x \bar x = q \bar q$, $e^+ e^-$, $\mu^+ \mu^-$,
are included in the analysis.
The expected signal events,
[$e^+ e^- \to HZ$, $Z \to x \bar x$, $H \to \tau^+ \tau^-$, $\tau^\pm \to \nu \pi^\pm$],
produced at the ILC and FCC-ee are 414 and 691, respectively.
At the ILC and (FCC-ee), after imposing the requirement, 
$|m_{\rm recoil} - m_H| < 5$ GeV,
385 (663) signal events survive.
We estimated that 20 (36) background events contribute
to this phase-space region. 
We perform 100 pseudo-experiments for each benchmark collider and estimate the statistical uncertainties on the measurements.

To take into account the energy mismeasurement, 
we smear the energies of all visible particles in the final state as
\be
E^{\rm true} \,\to\, E^{\rm obs} = (1 + \sigma_E \cdot \omega) \cdot E^{\rm true}
\ee
with the energy resolution $\sigma_E = 0.03$ \cite{Behnke:2013lya, FCC:2018evy}
for both ILC and FCC-ee,
where $\omega$ is a random number drawn from the normal distribution.

\subsection{Solving kinematical constraints}
\label{sec:kinematical}

\begin{table*}[t!]
\begin{center}
\begin{tabular}{|c || c | c|} 
 \hline
  & ~~~ILC~~~ & ~~~FCC-ee~~~ \\ 
 \hline \hline
 $C_{ij}$ & 
 $\begin{pmatrix} 
 -0.600 \pm 0.210 & 0.003 \pm 0.125 & 0.020 \pm 0.149 \\
 0.003 \pm 0.125 & -0.494 \pm 0.190 & 0.007 \pm 0.128 \\
 0.048 \pm 0.174 & 0.0007 \pm 0.156 & 0.487 \pm 0.193  
 \end{pmatrix}$
 & $\begin{pmatrix}
 -0.559 \pm 0.143 & -0.010 \pm 0.095 & -0.014 \pm 0.122 \\
 -0.010 \pm 0.095 & -0.494 \pm 0.139 & -0.002 \pm 0.111 \\
 0.012 \pm 0.124 & 0.020 \pm 0.105 & 0.434 \pm 0.134 
 \end{pmatrix}$  \\ 
 \hline
 $E_k$ & $-1.057 \pm 0.385  $ &  $ -0.977 \pm 0.264 $ \\
 \hline
 ${\cal C}[\rho]$  & $0.030\pm 0.071$ & $0.005\pm 0.023$  \\
 \hline
 ${\cal S}[\rho]$  & $1.148\pm 0.210 $ & $1.046\pm 0.163$  \\
 \hline
 $R_{\rm CHSH}^*$ & $0.769 \pm 0.189$ & $0.703 \pm 0.134$  \\
\hline
\end{tabular}
\end{center}
\caption{Result of quantum property measurements with a simple kinematical reconstruction method.
\label{tb:result_1}}
\end{table*}

Because of the presence of neutrinos in Eq.\ \eqref{eq:decay},
the momenta of two taus are not measured.
To perform measurements of the $C$-matrix and $R_{\rm CHSH}^*$, the momenta of two neutrinos must be reconstructed by solving kinematical constraints.  
For six unknown momentum components, there are two mass-shell constraints: $m_\tau^2 = (p_{\nu_\tau} + p_{\pi^-})^2$ and $m_\tau^2 = (p_{\bar \nu_\tau} + p_{\pi^+})^2$,
and four conditions from the energy-momentum conservation:
$(P_{\rm in} - p_Z)^\mu = (p_{\nu_\tau} + p_{\pi^-} + p_{\bar \nu_\tau} + p_{\pi^+})^\mu$.
By solving those six constraints for the six unknowns, an event can be fully reconstructed up to twofold solutions: $i_s = 1,2$ (see Appendix \ref{app:recon} for details).

The system is first boosted to the rest frame of $H$.
For each solution, $i_s$, we then boost the system to the reconstructed rest frame of $\tau^{-}$ and calculate the $r,n,k$ components of the $\pi^{-}$ direction, i.e.~$(u_r^{i_s}, u_n^{i_s}, u_k^{i_s})$.
In the same way, the $\pi^+$ direction, $(\bar u_r^{i_s}, \bar u_n^{i_s}, \bar u_k^{i_s})$, are obtained at the reconstructed rest frame of $\tau^+$.
We estimate the $C$-matrix elements with Eq.\ \eqref{eq:C_from_uu}.
For the Bell inequality test, $R^*_{\rm CHSH} \equiv R_{\rm CHSH}( {\bf a}_*, {\bf a'}_*, {\bf b}_*, {\bf b'}_*)$
is calculated using Eqs.~\eqref{eq:SCHSH_u} and \eqref{eq:max_abab}.
Both solutions, $i_s = 1,2$, are included 
in the calculation of $C_{ab}$ and $R^*_{\rm CHSH}$.

The result of the measurements for $C_{ab}$, $E_k$, ${\cal C}[\rho]$, ${\cal S}[\rho]$ and $R^*_{\rm CHSH}$ is summarised in Table \ref{tb:result_1}.
We see that the $C$-matrix is measured as a diagonal form with good accuracy.  
However, the diagonal elements are far off from the true values, $C = {\rm diag}(1,1,-1)$. 
Not only are the magnitudes significantly less than one, 
but also the signs are flipped for all diagonal components.
We also see no clear indication of the quantum correlations, i.e.\ entanglement ($E_k > 1$, ${\cal C}[\rho] > 0$), steerability (${\cal S}[\rho] > 1$)
and CHSH violation ($R^*_{\rm CHSH} > 1$). 

We identify two main reasons for this disappointing result.  
The first is the effect of false solutions of the kinematic reconstruction.  
The false solutions contribute to 
the measurements as much as the true solutions.\footnote{We however checked that
when smearing is turned off, even if only false solutions are used for the measurements, the true values for $C_{ab}$ (and therefore also for $R^*_{\rm CHSH}$ and $E_k$) are recovered as in the case where only true solutions are used.
When smearing is switched on,
both solutions are different from the MC truth 
and we therefore loose the notion of true and false solutions. 
}
The other effect is the smearing of the beam energies and the energy mismeasurements for the final state particles.
These impact the reconstruction of the tau momenta, in particular the direction of the tau leptons. 
In addition, since the tau leptons are highly boosted, a small error on their directions results in a large error on the 
$\pi^{\pm}$ distribution measured at  
the reconstructed  $\tau^{\pm}$ rest frame.

\subsection{Log-likelihood with the impact parameters}
\label{sec:likelihood}

\begin{table*}[t!]
\begin{center}
\begin{tabular}{|c || c | c|} 
 \hline
  & ~~~ILC~~~ & ~~~FCC-ee~~~ \\ 
 \hline \hline
 $C_{ij}$ & 
 $\begin{pmatrix} 
 0.830 \pm 0.176 & 0.020 \pm 0.146 & -0.019 \pm 0.159 \\
 -0.034 \pm 0.160 & 0.981 \pm 0.1527 & -0.029 \pm 0.156 \\
 -0.001\pm 0.158 & -0.021 \pm 0.155 & -0.729\pm 0.140
 \end{pmatrix}$
 & $\begin{pmatrix}
 0.925 \pm 0.109 & -0.011 \pm 0.110 & 0.038 \pm 0.095 \\
 -0.009 \pm 0.110  & 0.929 \pm 0.113  & 0.001 \pm 0.115  \\
 -0.026 \pm 0.122  & -0.019 \pm 0.110  & -0.879 \pm 0.098  
 \end{pmatrix}$  \\ 
 \hline
 $E_k$ & $2.567 \pm 0.279$ & $2.696  \pm 0.215$  \\
 \hline
 ${\cal C}[\rho]$  & $0.778\pm 0.126$ & $0.871 \pm 0.084$  \\
 \hline
 ${\cal S}[\rho]$  & $1.760 \pm 0.161 $ & $1.851 \pm 0.111$  \\
 \hline
 $R_{\rm CHSH}^*$ & $1.103 \pm 0.163$ & $1.276 \pm 0.094$  \\
 \hline
\end{tabular}
\end{center}
\caption{Result of quantum property measurements with a log-likelihood method incorporating the impact parameter information.
\label{tb:result_2}}
\end{table*}

We now discuss how to overcome the limitations identified in the previous section.
We note that the information obtained from the impact parameter measurements of tau decays has not been employed. 
Since tau leptons are marginally long-lived, $c \tau = 87.11$ 
$\mu$m \cite{Workman:2022ynf},
and highly boosted, one can observe a mismatch 
between the interaction point and the origin of the $\pi^\pm$ in $\tau^\pm \to \nu \pi^\pm$.
The impact parameter $\vec{b}_\pm$ is the minimal displacement of the extrapolated $\pi^\pm$ trajectory from the interaction point. 
The magnitude of the impact parameter $|\vec b_{\pm}|$ follows an exponentially falling distribution with the mean $|\vec b_{\pm}| \sim 100 \, \mu$m for $E_{\tau^\pm} \sim m_H/2$,
which is significantly larger than the experimental resolutions \cite{Behnke:2013lya}.
In our numerical simulation, 
we take constant values 
$\sigma_{b_T} = 2 \, \m$m (transverse)
and $\sigma_{b_z} = 5 \, \m$m (longitudinal) 
for the impact parameter resolutions,
although the actual resolutions are 
functions of the track momentum and
the polar angle $\theta^*$ from the beam direction.
The above modeling with the constant parameters gives a reasonable approximation for
the track momentum $\sim 100$ GeV
and $\theta^* \gtrsim 20\degree$ 
as can be seen in Figure II-3.10 in ref.\ 
\cite{Behnke:2013lya}.

If all quantities are accurately measured, the impact parameter, ${\vec{b}_\pm}$, from the $\tau^\pm \to \nu \pi^\pm$ decay,
is related to the directions of $\taup$ and $\pip$
and their angle $\Theta_\pm$ by \cite{Hagiwara:2016zqz}
\ba
\vec b_\pm &=& |\vec b_\pm| \cdot \left[ {\bf e}_\taupm  \cdot \sin^{-1} \Theta_\pm 
\,-\, {\bf e}_\pipm \cdot \tan^{-1} \Theta_\pm   \right]
\nonumber \\
&\equiv& \vec b_\pm^{\rm reco} \left( {\bf e}_\taupm \right)
\,,
\label{eq:b}
\ea
where ${\bf e}_\taupm$ and ${\bf e}_\pipm$ 
are the unit vectors pointing to the directions of $\tau^\pm$ and $\pi^\pm$, respectively, and $\cos \Theta_\pm \equiv ({\bf e}_\taupm \cdot {\bf e}_\pipm)$.
In the second line, we defined a 3-vector function
$\vec b_\pm^{\rm reco} \left( {\bf e}_\taupm  \right)$ and emphasised its dependence on ${\bf e}_\taupm$.

We use this information to  curb the effects of energy mismeasurement.
First, we shift the energy of a visible particle $\alpha$ ($\a = \pi^\pm, x, \bar x$) from the observed value as
\be
E^{\rm obs}_{\a} \to E_{\a}(\delta_\a) = (1 + \sigma_E \cdot \delta_\a) \cdot E_{\a}^{\rm obs}\,,
\ee
where $\delta_\a$ is a nuisance parameter characterising the amount of the shift with respect to the energy resolution $\sigma_E$.
Using these shifted energies, we solve the kinematical constraints, as outlined in Appendix \ref{app:recon}, and obtain the tau directions as functions of the nuisance parameters, ${\bf e}^{i_s}_{\taupm}( \pmb{\delta} )$, up to twofold solutions, $i_s = 1,2$,
where $\pmb{\delta} = \{ \delta_\pi^+, \delta_\pi^-, \delta_x, \delta_{\bar x} \}$.
Based on the mismatch between the {\it observed}
and {\it reconstructed} impact parameters,
\be
\vec \Delta^{i_s}_{b_\pm}( \pmb{\delta} ) 
\,\equiv\, 
\vec b_\pm - \vec b_\pm^{\rm reco} \left( {\bf e}^{i_s}_{\taup} ( \pmb{\delta} )
\right)\,,
\ee
we define a contribution to the log-likelihood for a solution $i_s$ as
\be
L^{i_s}(\pmb{ \delta }) \,=\, L^{i_s}_{+}(\pmb{ \delta }) + L^{i_s}_{-}(\pmb{ \delta })
\ee
with
\be
L^{i_s}_{\pm}(\pmb{\delta}) \,=\, \frac{ [\Delta^{i_s}_{b_\pm}( \pmb{\delta} )]_x^2 + [\Delta^{i_s}_{b_\pm}( \pmb{\delta} )]_y^2 }{ \sigma^2_{b_T} } \,+\,
\frac{ [\Delta^{i_s}_{b_\pm}( \pmb{\delta} )]_z^2 }{ \sigma^2_{b_z} }\,.
\ee
The total log-likelihood function is then defined as
\be
L(\pmb{\delta}) \,=\, {\rm min} \left[ L^1( \pmb{\delta} ), L^2( \pmb{\delta} ) \right] \,+\, \delta_{\pi^+}^2\,+\, \delta_{\pi^-}^2\,+\, \delta_{x}
^2\,+\, \delta_{\bar x}^2 \,.
\ee

The log-likelihood function, $L(\pmb{ \delta })$, is to  be minimised over the nuisance parameters, $\pmb{ \delta }$.
We denote the location of the minimum by $\pmb{ \delta}^* $.
We define ``the most likely'' solution $i_*$ as the solution that gives the smaller $L^{i_s}$, i.e.\
$
L^{i_*}( \pmb{ \delta}^* ) \,=\,
{\rm min} \left[ L^1( \pmb{ \delta}^*), L^2(\pmb{ \delta}^*) \right]\,.
$
Our best guess for the tau lepton momenta is, therefore, given by 
\be
p_{\taupm}^* = p^{i_*}_{\taupm}(\pmb{ \delta}^* )\,.
\ee
In what follows we use $p_{\taupm}^*$ in the quantum property measurements. 

In Table \ref{tb:result_2} we show the result of our quantum property measurements when the impact parameter information 
of tau decays
is incorporated in the log-likelihood. 
We see that for both ILC and FCC-ee the components of the $C$-matrix are correctly measured including the sign.  
The entanglement signature $E_k$ 
and the concurrence ${\cal C}[\rho]$
are also measured with a good accuracy 
and the formation of entanglement 
($E_k > 1$ and ${\cal C}[\rho] > 0$)
is observed at
more  than 5\,$\sigma$.
The steerability variable, ${\cal S}[\rho]$,
is also well measured and the Standard Model value, ${\cal S}[\rho] = 2$, is more or less reproduced. 
The steerability condition, ${\cal S}[\rho] > 1$, is observed at $\sim 4\,\sigma$ 
for the ILC and $\gg 5\,\sigma$ for 
the FCC-ee.
Observation of Bell-nonlocalilty is the most challenging one since it is the strongest quantum correlation.
As can be seen in the last line in Table \ref{tb:result_2},
the violation of the CHSH inequality is confirmed at the FCC-ee at $\sim 3\,\sigma$ level,
while $R^*_{\rm CHSH} > 1$ is not observed at the ILC beyond the statistical uncertainty.
The superior performance of FCC-ee is attributed to the fact that the beam energy resolution of FCC-ee is much better than ILC.
The precise knowledge of the initial state momentum is crucial to accurately reconstruct the rest frame of $\tau^\pm$.

\section{CP measurements}
\label{sec:CP}

Since the $C$-matrix is sensitive to the CP phase $\delta$,
one can use the result of $C$-matrix measurement and derive a constraint on $\delta$. 
From Eq.\ \eqref{eq:Cmat} we see that
only the $rn$ part (i.e.\ the upper-left $2 \times 2$ part) of the $C$-matrix is sensitive to $\delta$. By comparing the measured $C$-matrix entries in the $rn$ part 
and the prediction in Eq.\ \eqref{eq:Cmat}, we construct the $\chi^2$ function as
\ba
\chi^2(\delta) &=& \frac{ \left( C_{rr} - \cos 2 \delta \right)^2 }{ \sigma^2_{rr} } +  
\frac{ \left( C_{rn} - \sin 2 \delta \right)^2 }{ \sigma^2_{rn} } 
\nonumber \\
&& +\, \frac{ \left( C_{nn} - \cos 2 \delta \right)^2 }{ \sigma^2_{nn} } 
+ \frac{ \left( C_{nr} + \sin 2 \delta \right)^2 }{ \sigma^2_{nr} } \,,
\label{eq:chi2_delta}
\ea
where $C_{ij}$ and $\sigma_{ij}$ are the central value and the standard deviation, respectively, obtained from the analysis in subsection \ref{sec:likelihood}. The goodness of fits are found to be
$\chi^2_{\rm min}({\rm ILC})/{\rm d.o.f.} = 0.93/3$ 
and
$\chi^2_{\rm min}$(FCC-ee)/{\rm d.o.f.}$ = 0.86/3$
for each benchmark collider.

\begin{table}[t!]
\begin{center}
\begin{tabular}{ c | c | c } 
~~~~CL~~~~ & ILC & FCC-ee \\ [0.5ex] 
\hline 
 68.3\,\% & $[-7.94\degree,6.20\degree]$  & $[-5.17\degree, 5.11\degree]$ \\
 95.5\,\% & $[-10.89\degree,9.21\degree]$  & $[-7.36\degree, 7.31\degree]$ \\
 99.7\,\% & $[-13.84\degree,12.10\degree]$  & $[-9.21\degree, 9.21\degree]$   \\
\end{tabular}
\end{center}
\caption{Expected sensitivities on the CP phase $\delta$.
\label{tb:cp}}
\end{table}

\begin{figure}[t!]
    \centering
	\includegraphics[width=0.95\linewidth]{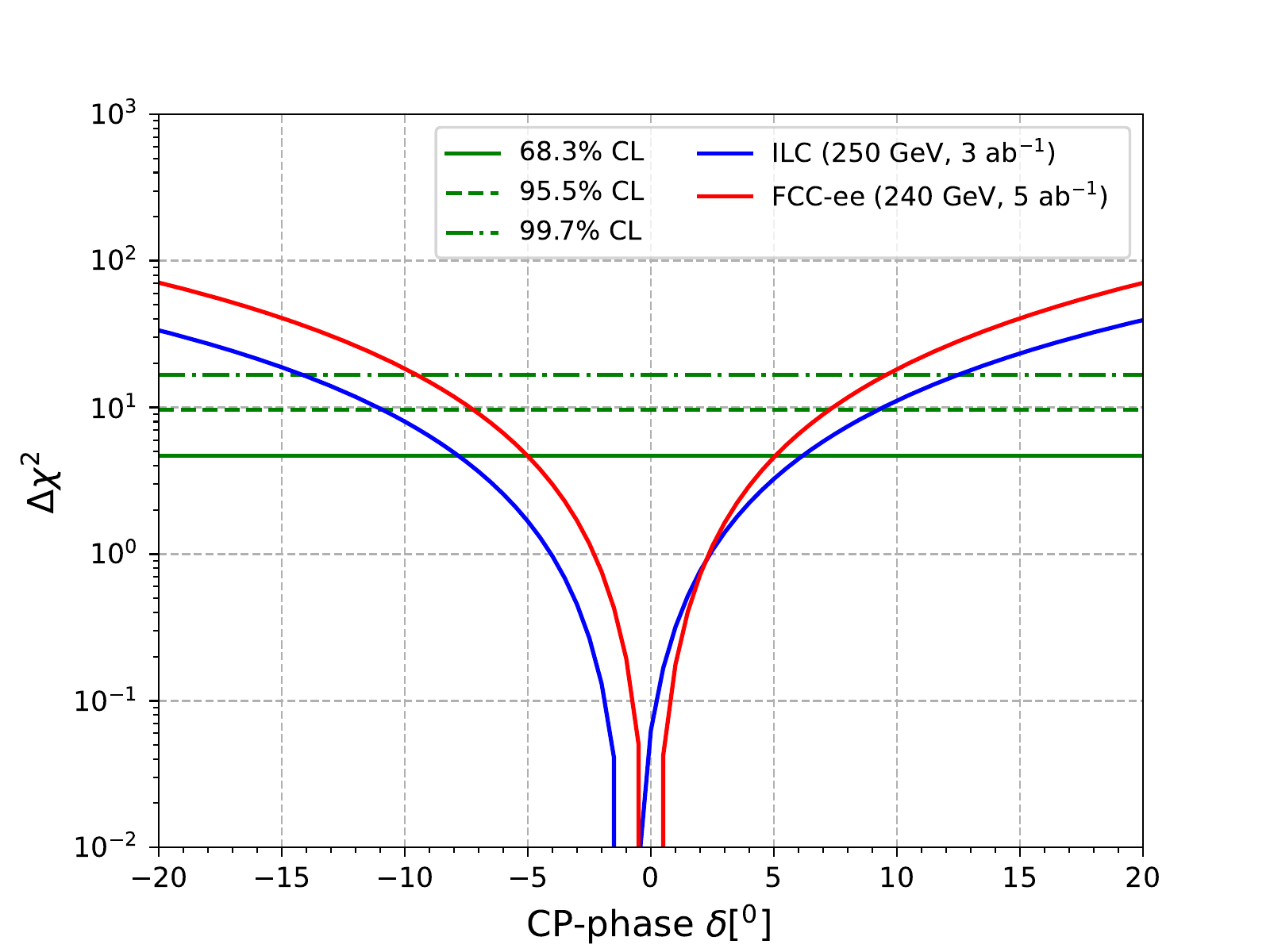}\,
	\caption{$\Delta \chi^2$ as a fuction of the CP phase $\delta$.
	}
	\label{fig:chi2}
\end{figure}

The minimum of $\chi^2$ appears at the vicinity of three CP-conserving points: $\delta = 0$, $\pm 180\degree$ (CP-even) and $\pm 90\degree$ (CP-odd).
Focusing on the minimum around $\delta = 0$,
the 1, 2 and 3 $\sigma$ regions of $\delta$
obtained from this analysis are listed in Table \ref{tb:cp}.
The analysis is based on 
$\Delta \chi^2(\delta) \equiv \chi^2(\delta) - \chi^2_{\rm min}$, whose values around $\delta = 0$ are plotted in Fig.~\ref{fig:chi2}.
We note that the allowed windows are asymmetric. This is due to the statistical uncertainty 
of the 100 pseudo-experiments.

We see that the resolution of $\delta$  
obtained from this analysis 
is roughly $\sim 7.5\degree$ (ILC) and $\sim 5\degree$ (FCC-ee) at 1 $\sigma$ level.
These results should be compared with the resolutions 
obtained in the standard approach for the $H \to \tau^+ \tau^-, \tau^\pm \to \pi^\pm \nu$ channel,
which exploits the angle $\varphi^\star$ 
\cite{Kramer:1993jn,Was:2002gv}
defined 
between the two planes, each  
spanned by the pair of momentum 3-vectors 
$(\vec{p}_{\pi^+}, \vec{p}_{\tau^+})$
and
$(\vec{p}_{\pi^-}, \vec{p}_{\tau^-})$
in the Higgs rest frame.
Using the same event reconstruction technique  
described in section \ref{sec:likelihood}
and the statistical method based on 100 pseudo-experiments,
we find the resolution of the CP-phase 
with the $\varphi^\star$ method
is $6.4\degree$ for FCC-ee.
This shows that the proposed method based on Eq.\ \eqref{eq:chi2_delta} is at least as good as the standard method with $\varphi^\star$.

In the literature, the $H \tau \tau$ CP phase measurement 
at ILC using different tau decay modes
have also been explored \cite{Bower:2002zx,Desch:2003rw}.
The studies in \cite{Harnik:2013aja, Jeans:2018anq}
exploit the $\tau^\pm \to \rho^\pm \nu$ channel 
and suggest that the sensitivity can reach $\sim 4 \degree$,\footnote{Note, however, that these analyses do not include the effect of energy mismeasurement.}
which is in line with the expectations of first theoretical studies, i.e.,  $2.8\degree$ \cite{Berge:2013jra}, including 
also other decay modes, e.g.\ $\tau^\pm \to a^\pm \nu$ and $\tau^\pm \to \ell^\pm \nu \nu$.
A recent study \cite{Chen:2017bff} using a likelihood analysis based on the Matrix-Element
claims that the CP-phase can be measured in the accuracy of 
$2.9\degree$ 
at the ILC.

At HL-LHC, the resolution of the $H \tau\tau$ CP-phase 
measurement is expected to reach  
$\sim 11\degree$
using the tau decay modes
$\tau^\pm \to \pi^\pm \nu$ \cite{Hagiwara:2016zqz}
and $\tau^\pm \to \rho^\pm \nu$ \cite{Harnik:2013aja}.
Combining comprehensive decay modes with the Matrix-Element
likelihood method, the resolution may reach
$5.2\degree$ \cite{Chen:2017nxp}.
On the other hand, however, ref.\ \cite{Askew:2015mda}
claims 
that the detector effect severely impacts on 
the performance of CP measurements at HL-LHC
and the CP phase hypothesis $\delta = 0$ can be distinguished from $\delta = 90\degree$ 
only at 95\,\% CL with $\tau^\pm \to \pi^\pm \nu$ channel.

\subsection*{Model independent CP test}

Under the CP conjugation, the $C$-matrix transforms as 
$C \xrightarrow{CP} C^T$.
This fact can be used for a model-independent test for CP violation.
To measure the asymmetry in the $C$-matrix, 
we define
\be
A = (C_{rn} - C_{nr})^2 + (C_{nk} - C_{kn})^2 + (C_{kr} - C_{rk})^2 \geq 0 \,.
\ee
An experimental verification of $A \neq 0$ immediately confirms violation of CP. 

From the analysis 
described in subsection \ref{sec:likelihood},
$A$ is measured as:
\be
A \,=\,
\left\{
\begin{array}{ll}
0.168 \pm 0.131 &~~~{\rm (ILC)}
\\
0.081 \pm 0.060 &~~~({\rm FCC}\text{-}{\rm ee})
\end{array}
\right.
\ee
Here, the error corresponds to a 1\,$\sigma$ statistical uncertainty obtained from 100 pseudo-experiments. 
The result is consistent with 
the Standard Model (i.e.\ absence of CP violation)
at $\sim 1\,\sigma$ level.

In the explicit model defined by Eq.\ \eqref{eq:Htautau},
we have set $A = 4 \sin^2(2 \delta)$.
One can interpret  
the above model-independent result 
within this model and derive bounds on $\delta$.
In the domain around $\delta = 0$, the following limits are obtained at $1\,\sigma$:
\be
|\delta| <
\left\{
\begin{array}{l}
7.9\degree ~~~{\rm (ILC)}
\\
5.4\degree~~~({\rm FCC}\text{-}{\rm ee})
\end{array}
\right.\,,
\ee
consistently with the limits obtained in the $\chi^2$ analysis (see Table \ref{tb:cp}).


\section{Conclusions}
\label{sec:Conclusions}

We have investigated the feasibility of testing various quantum properties, such as entanglement, steering and Bell-nonlocality, 
with the tau spin pairs in $H \to \tau^+ \tau^-$
at the future $e^+ e^-$ colliders.
Two collider benchmark scenarios, ILC and FCC-ee, have been considered with the parameters listed in Table \ref{tb:colliders}.
We found that although the tau spin pairs in 
$H \to \tau^+ \tau^-$ are maximally entangled (${\cal C}[\rho] = 1$) and saturate the upper bounds of the steering and Bell-nonlocality measures (${\cal S}[\rho] = 2$ and $R^{\rm max}_{\rm CHSH} = \sqrt{2}$), experimental observation of those quantum properties is non-trivial since 
quantum correlation is easily smeared away once the detector and beam energy resolution are taken into account (see Table \ref{tb:result_1}). 
In order to curb the effect of energy mismeasurements, 
we developed a log-likelihood method by measuring and utilising the consistency between the reconstructed tau momenta and the measured direction of impact parameters of tau decays.

\begin{table}[t!]
\begin{center}
\begin{tabular}{ c | c | c } 
 & ~~~ILC~~~ & ~~FCC-ee~~ \\ [0.5ex] 
\hline 
~~Entanglement~~  & $> 5\sigma$  & $\gg 5\sigma$  \\
~~Steerability~~  & $\sim 4\sigma$  & $\gg5\sigma$  \\
~~Bell-nonlocality~~  & --  & $\sim 3\sigma$  \\
~~CP-phase $\delta$ ($\Delta \chi^2$)~~  & $[-7.94\degree,6.20\degree]$  & $[-5.17\degree,5.11\degree]$ \\
~~CP-phase $|\delta|$ (A)~~  & $< 7.9\degree$  & $< 5.4\degree$ 
\end{tabular}
\end{center}
\caption{
Expected performance of the quantum property measurements.
The last two lines indicate the 1\,$\sigma$
resolution for 
the CP phase of the $H \tau \tau$ coupling obtained from the $\Delta \chi^2$
analysis and the asymmetry ($A$) measurement.
\label{tb:summary}}
\end{table}

Using MC simulations of 100 pseudo-experiments (for each ILC and FCC-ee), we have demonstrated that accurate quantum property measurements are possible at ILC and FCC-ee, including the effects
of the detector and beam energy resolution.
Our main result is summarised in Table \ref{tb:result_2}.
Table \ref{tb:summary}  summarises our results by showing the statistical significance for observation of the quantum properties: entanglement, steerability and Bell-nonlocality.
Our analysis shows that violation of the CHSH inequality  cannot be observed even at 1\,$\sigma$ level at ILC due to large beam energy resolutions, while it can be observed at 3\,$\sigma$ level at FCC-ee.

The spin correlation of tau pairs in $H \to \tau^+ \tau^-$ is sensitive to the CP-phase of the $H \tau \tau$ interaction. 
We have proposed a model-independent test of CP violation with the measurement of the spin correlation matrix.
We found this method can constrain the CP-phase of the $H \tau \tau$ interaction up to 7.9$\degree$ and 5.4$\degree$ at ILC and FCC-ee, respectively, at 1\,$\sigma$ level (see Table \ref{tb:summary}), similarly to dedicated analyses at the corresponding colliders. 

Finally, we comment on a subtlety of the collider test of Bell-nonlocality \cite{Abel:1992kz}.
In collider experiments, the spin of particles is not directly measured but only inferred from angular distributions of their decay products. 
For example, in our analysis we calculated $R_{\rm CHSH}$ with Eq.\ \eqref{eq:angle-to-spin}, which relates the spin correlation to the angular correlation.
The problem is this relation is based on Eq.\ \eqref{eq:prob}, which is derived using quantum mechanics, quantum field theory in particular.
In fact, one could think of a class of local hidden variable (LHV) theories that predict the pion directions directly through a set of hidden variables as 
${\bf u}(\lambda)$
and ${\bf \overline u}(\lambda)$.
In that case, analogously to Eq.\ \eqref{eq:R},
Bell's argument derives a CHSH inequality directly about the pion directions:
\be
\widetilde R_{\rm CHSH} \equiv \frac{1}{2} 
\left| \bra u_a \bar u_b \ket    
- \bra u_a \bar u_{b'} \ket 
+ \bra u_{a'} \bar u_b \ket 
+ \bra u_{a'} \bar u_{b'} \ket \right|
\, \leq \, 1\,.
\label{eq:R2}
\ee
Although this inequality is more general and applicable for any LHV theories, it is too weak.
Since $\widetilde R_{\rm CHSH} = \frac{| \a_{f,d} \a_{f',d'}|}{9}  R_{\rm CHSH}$
and $R_{\rm CHSH} \leq \sqrt{2}$ in quantum mechanics,
the inequality \eqref{eq:R2} is not violated even in quantum mechanics and cannot be used to falsify LHV theories.

Having said that, the angular distribution \eqref{eq:prob} is well tested elsewhere experimentally.
As long as one believes the physical picture in which the tau lepton is a spin-{1/2} particle and its decay products obey the angular distribution \eqref{eq:prob}, observation of $R_{\rm CHSH} > 1$ excludes the LHV theories that try to explain the spin correlation of tau pairs in $H \to \tau^+ \tau^-$.


\section*{Acknowledgements}
K.S.\ thanks Zbigniew Was and Elzbieta Richter-Was for valuable discussions.
K.S.\ also thanks Iwate University for a visit.
P.L.\ are partially supported by the National Science Centre, Poland, under research grant 2017/26/E/ST2/00135. The work of M.M.A.\ was supported by the French Agence Nationale de la Recherche (ANR) grant no.\ ANR-21-CE31-0023 (PRCI SLDNP) and by the National Science Centre, Poland, under research grant 2017/26/E/ST2/00135.
The works of K.M are supported in part by JSPS KAKENHI Grant No. 20H05239, 21H01077 and 21K03583.
K.S.\ is partially supported by the National Science Centre, Poland, under research grant 2017/26/E/ST2/00135 and the Norwegian Financial Mechanism for years 2014-2021, grant DEC-2019/34/H/ST2/00707. F.M. is partially supported by the F.R.S.- FNRS under the “Excellence of Science” EOS be.h project no. 30820817.


\appendix

\section{An operational definition of steerability}\label{ap:steering}

In this section we provide an operational definition of steerability
\cite{Jones2007,PhysRevLett.98.140402}
and derive the mathematical definition given in the main text 
(Eqs.\ \eqref{eq:steer1} and \eqref{eq:steer2}) from it.

Consider the following experiment.
Prior to the experiment, 
Alice and Bob agree that 
Alice prepares a biparticle quantum state, $\rho$, keeps one particle with her and sends the other one to Bob.
Bob's task is to prove, without trusting Alice, that Alice can ``steer'' (change) the state of his particle
by her measurement.
He carries out his task by asking Alice to perform some measurement on her particle and report the outcome via a classical communication.  
He can also make a measurement on his particle.
Bob can repeat the process any number of times.
If he succeeds the task, the state $\rho$ is said to be
steerable by Alice.

Without any information from Alice, the local state of Bob's particle is 
\be
\rho_B = {\rm Tr}_A ( \rho  )\,,
\label{eq:rB}
\ee
where ${\rm Tr}_A$ is the partial trace for Alice's particle.  
Let ${\cal M}_A$ and ${\cal M}_B$ be the sets of all possible observables of Alice and Bob, respectively.
If Alice measures an observable ${\cal A} \in {\cal M}_A$, 
the probability of observing the outcome $a$ is
\be
q_a^{\cal A} = {\rm  Tr}[F_a^{\cal A} \rho \,  (F_a^{\cal A})^\dagger]\,,
\ee
where $F_a^{\cal A}$ is Alice's Positive Operator-Valued Measure (POVM) with $F_a^{\cal A} \geq 0$ and
$\sum_a (F_a^{\cal A})^\dagger F_a^{\cal A} = 1$.
For projective measurements (closed systems), $F_a^{\cal A} = | a \ket \bra a |$,
with ${\cal A} | a \ket = a | a \ket$.
Because of Alice's measurement, the quantum state collapses into the post-measurement state as
\be
\rho ~\xrightarrow{a,{\cal A}}~ 
\rho^{\rm post}_{a,{\cal A}}
\,=\, \frac{ F_a^{\cal A} \rho \,  (F_a^{\cal A})^\dagger  }{ q^{\cal A}_a }\,.
\ee
After this event, Bob's local state becomes
\be
\rho^{a,{\cal A}}_B = {\rm Tr}_A( \rho_{a,{\cal A}}^{\rm post} )\,.
\ee

To test whether Alice can indeed steer Bob's state, 
Bob would first check if she reports the outcome $a$ with the correct frequency $q^{\cal A}_a$, which he can calculate from $\rho$. 
Second, he makes measurements on his particles (one measurement at each time, but he makes many different measurements in different times) and checks if his state is indeed changed from $\rho_B$ to $\rho_B^{{\cal A}, a}$ when she measures ${\cal A}$ and reports $a$.
Bob does these checks for all possible observables of Alice.
If all of these checks pass, Bob would be inclined to agree that Alice can steer Bob's particle. 
However, he must consider the following ``cheating'' scenario.

In the cheating scenario, Alice sends Bob a one-particle local state $\rho_B^\lambda$, parameterised by some variables $\lambda$ with  probability distribution $P(\lambda)$,  such that $\rho_B = \sum_\lambda P(\lambda) \rho_B^\lambda$.
Alice has the full information of $\lambda$ and $\rho_B^\lambda$ every time she sends it.
If Bob asks Alice to measure ${\cal A}$, Alice will tell him the outcome is $a$ with some frequency $p(a|{\cal A},\lambda)$, depending on $\lambda$.
Bob's first check will pass, if this function satisfies 
\be
\sum_\lambda p(a|{\cal A},\lambda) P(\lambda) = q_a^{\cal A}  \,.
\label{eq:Qqx}
\ee

For his second check, he would collect all the states in which Alice reported the outcome $a$ in her ${\cal A}$ measurement. 
Then, he makes measurements on this correction to check if it is indeed $\rho_B^{a,{\cal A}}$.
In the above scenario, he has a local state $\rho_B^\lambda$
with the probability $p(a|{\cal A},\lambda) P(\lambda)/q^{\cal A}_a$,
so this collection is a mixed state $\sum_\lambda p(a|{\cal A},\lambda) P(\lambda) \rho_B^\lambda/q^{\cal A}_a$.
Therefore, Bob's second check would pass
if $p(a|{\cal A},\lambda)$ satisfies 
\be
\frac{1}{q^{\cal A}_a} \sum_\lambda p(a|{\cal A},\lambda) P(\lambda) \rho_B^\lambda = \rho_B^{a,{\cal A}}\,.
\label{eq:QqB}
\ee

The above consideration tells that if there exists a function $p(a|{\cal A},\lambda)$ satisfying both Eqs.\ \eqref{eq:Qqx} and \eqref{eq:QqB} for all ${\cal A} \in {\cal M}_A$,
Bob cannot exclude the possibility that Alice is cheating.
Conversely, if such functions do not exist, Bob must conclude Alice can indeed steer the local state of his particle. 
Since $q_a^{\cal A}$ and $\rho_B^{a,{\cal A}}$ depend on $\rho$
and the statement is about all ${\cal A} \in {\cal M}_A$,
Alice's steerability depends only on $\rho$.

In the cheating scenario, Bob's local state after the Alice's measurement $(a, {\cal A})$ is given by 
Eq.\ \eqref{eq:QqB}.
If he measures ${\cal B} \in {\cal M}_B$ on his particle, the probability of obtaining the outcome $b$ is 
\be
{\rm Tr} [ \rho_B^{a,{\cal A}} F_b^{\cal B} ]
= \frac{1}{q^{\cal A}_a} \sum_\lambda p(a|{\cal A}, \lambda) P(\lambda) {\rm Tr} [ \rho^\lambda_B F_b^{\cal B}]\,,
\ee
where 
$F_b^{\cal B}$ is Bob's POVM.
At the same time, the probability that Alice reports $a$ when Bob asks her to measure ${\cal A}$ is $q_a^{\cal A}$.
The joint conditional probability 
under which Alice and Bob measure ${\cal A}$ and ${\cal B}$, respectively, and then Alice reports $a$ and Bob finds $b$ is the product of these two probabilities, i.e.
\be
p(a,b|{\cal A},{\cal B}) = \sum_\lambda P(\lambda) p(a|{\cal A},\lambda) {\rm Tr}[\rho_B^\lambda F_b^{\cal B}]\,.
\label{eq:steerability_cond}
\ee
This means that the definition of steerability can also be phrased in the following way:
\emph{the state $\rho$ is steerable by Alice if there does not exist a set of functions $p(a|{\cal A},\lambda)$ and one-particle local states $\rho_B^\lambda$, such that the joint conditional probability $p(a,b|{\cal A},{\cal B})$ is described by Eq.\ \eqref{eq:steerability_cond} 
for all ${\cal A} \in {\cal M}_A$ and ${\cal B} \in {\cal M}_B$.
}
This agrees with the mathematical definition provided in the main text 
(Eqs.\ \eqref{eq:steer1} and \eqref{eq:steer2}).

\section{Spin vs angular correlations}\label{ap:correlation}

The spin correlation $\bra s_a \bar s_b \ket$ of $\tau^- \tau^+$
and the angular correlation $\bra u_a \bar u_b \ket$ between the 
$\tau^- \tau^+$ decay products are related by Eq.\ \eqref{eq:angle-to-spin}.
To derive this result, we start by recalling 
Eq.\ \eqref{eq:prob}, i.e.\
 the conditional probability that the decay product, $d$, 
takes the direction ${\bf u}$ (at the rest frame of $\tau^-$) , when the tau spin is polarised into $\bf s$ direction, is given by
$$P( {\bf u} | {\bf s} ) \,=\, 1 + \alpha_{f,d} \, {\bf u} \cdot {\bf s}$$
with the normalisation 
$\int \frac{d \Omega_{\bf u}}{4 \pi} P( {\bf u} | {\bf s} ) = 1$.

We introduce the join probability that $\tau^-$ and $\tau^+$ are polarised into ${\bf s}$ and ${\bf \bar s}$,
and write it as $P({\bf s}, {\bf \bar s})$ with normalisation 
$\int \frac{d \Omega_{\bf s}}{4 \pi} \frac{d \Omega_{\bf \bar s}}{4 \pi} P({\bf s}, {\bf \bar s}) = 1$.
For arbitrary unit vectors ${\bf a}$ and ${\bf b}$,
the correlation between the $\tau^-$ and $\tau^+$ spin components, $s_a \equiv {\bf a} \cdot {\bf s}$ and $\bar s_b \equiv {\bf b} \cdot {\bf \bar s}$, can be written as
\be
\bra s_a \bar s_b \ket 
\,=\,
\int 
\frac{d \Omega_{\bf s}}{4 \pi} \frac{d \Omega_{\bf \bar s}}{4 \pi} 
({\bf a} \cdot {\bf s}) ({\bf b} \cdot {\bf \bar s})
P({\bf s},{\bf \bar s} )\,.
\label{eq:ss}
\ee
Similarly, the correlation between the components of the ${\bf u}$ and ${\bf \bar u}$ vectors is given by
\ba
\bra u_a \bar u_b \ket &=&
\int \frac{d \Omega_{\bf u}}{4 \pi} \frac{d \Omega_{\bf \bar u}}{4 \pi}
\frac{d \Omega_{\bf s}}{4 \pi} \frac{d \Omega_{\bf \bar s}}{4 \pi}
({\bf a} \cdot {\bf u}) ({\bf b} \cdot {\bf \bar u})
\nonumber \\
&&~~~\times\,
P( {\bf u} | {\bf s} )
P( {\bf \bar u} | {\bf \bar s} )
P( {\bf s},{\bf \bar s} )\,.
\label{eq:uaub_integral}
\ea

We carry out the integration $d \Omega_{\bf u}$
by expressing ${\bf u}$ in a polar coordinate 
where the pole is taken into the ${\bf s}$ direction
(we call this the $z$ direction).
Similarly, we represent  ${\bf \bar u}$
in a polar coordinate with the pole in
the ${\bf \bar s}$ direction ($z'$ direction).
Using these two coordinate systems, we have
\ba
&& {\bf u} \cdot {\bf s} \,=\, c_\theta,~~~
{\bf \bar u} \cdot {\bf \bar s} \,=\, c_{\theta'},~~~
\nonumber \\
&&{\bf a} \cdot {\bf u} = 
a_x s_\theta c_\phi + a_y s_\theta s_\phi + a_z c_\theta,
\nonumber \\
&& {\bf b} \cdot {\bf \bar u} = 
b_{x'} s_{\theta'} c_{\phi'} + b_{y'} s_{\theta'} s_{\phi'} + b_{z'} c_{\theta'},
\nonumber \\
&&a_z \,=\, {\bf a} \cdot {\bf s} \,=\, s_a,~~~
b_{z'} \,=\, {\bf b} \cdot {\bf \bar s}' \,=\, \bar s_b \,,
\ea
and Eq.\ \eqref{eq:uaub_integral} is expressed as
\ba
\bra u_a \bar u_b \ket &=&
\int \frac{ d c_{\theta} d \phi}{4 \pi} \frac{d c_{\theta'} d \phi'}{4 \pi}
\frac{d \Omega_{\bf s}}{4 \pi} \frac{d \Omega_{\bf \bar s}}{4 \pi}
\nonumber \\
&&
(a_x s_\theta c_\phi + a_y s_\theta s_\phi + a_z c_\theta)
\nonumber \\
&&
(b_{x'} s_{\theta'} c_{\phi'} + b_{y'} s_{\theta'} s_{\phi'} + b_{z'} c_{\theta'})
\nonumber \\
&&
(1 + \alpha_{f,d} c_\theta)(1 + \alpha_{f',d'} c_{\theta'})
P( {\bf s},{\bf \bar s} ) \,.
\ea
Any terms depending on $\phi$ or $\phi'$ will drop out 
by performing $d \phi$ and $d \phi'$ integrals, respectively.
The remainder is
\ba
\bra u_a \bar u_b \ket &=&
\int \frac{ d c_{\theta} }{2} \frac{d c_{\theta'} }{2}
\frac{d \Omega_{\bf s}}{4 \pi} \frac{d \Omega_{\bf \bar s}}{4 \pi}
\nonumber \\
&&
a_z c_\theta b_{z'} c_{\theta'}
(1 + \alpha_{f,d} c_\theta)
(1 - \alpha_{f',d'} c_{\theta'})
P( {\bf s},{\bf \bar s} ) 
\nonumber \\
&=&
\int
\left( \frac{d \Omega_{\bf s}}{4 \pi} \frac{d \Omega_{\bf \bar s}}{4 \pi}
s_a \bar s_b P( {\bf s},{\bf \bar s} ) 
\right)
\nonumber \\
&\times&
\Big(
\int \frac{d c_{\theta}}{2} \frac{d c_{\theta'}}{2} 
c_\theta c_{\theta'} (1 + \alpha_{f,d} c_{\theta})(1 + \alpha_{f',d'} c_{\theta'} )
\Big)\,,
\nonumber \\
\ea
where the first bracket on the RHS
is nothing but 
$\bra s_a \bar s_b \ket$ in Eq.\ \eqref{eq:ss}.
The second bracket produces $\frac{\alpha_{f,d} \alpha_{f',d'}}{9}$ and one obtains 
the result \cite{Abel:1992kz,ShionChen}
$$
\bra u_a \bar u_b \ket \,=\, \frac{\alpha_{f,d} \alpha_{f',d'}}{9} 
\bra s_a \bar s_b \ket \,.
$$

\section{Event reconstruction}\label{app:recon}

Since neutrinos are invisible in the detector, one has to reconstruct the neutrino momenta (or equivalently, the tau momenta) by solving kinematical constraints.
The 4-momentum of the initial $e^+ e^-$ pair, $P_{\rm in}^\mu$,
and the $Z$-boson, $p_Z^\mu$, are relatively accurately measured.  This motivate us to write the Higgs momentum as
\be
p_H^\mu = P_{\rm in}^\mu - p_Z^\mu \,.
\ee
The tau momenta, $\ptaup$ and $\ptaum$, are unknown but the sum is constrained by 
\be
\ptaup + \ptaum = p_H^\mu \,.
\label{eq:ptau_sum}
\ee
Each tau momentum is a 4-vector, so they can be expanded by four independent 4-vectors.  
We choose $p_H^\mu$, $p_{\pip}^\mu$, $p_\pim^\mu$ and $q^\mu$
as the basis vectors (neither orthogonal nor normalised), where we introduced
\be
q^\mu \,\equiv\, \frac{1}{m_H^2} \epsilon^{\m \n \r \s} p_H^\n \, p_\pip^r \, p_\pim^s \,,
\ee
which is orthogonal to the other basis vectors, $(q \cdot p_h) = (q \cdot \ptaup) = (q \cdot \ptaum) = 0$.
In terms of these basis vectors, the tau momenta are expanded as \cite{Harland-Lang:2011mlc, Harland-Lang:2012zen, Harland-Lang:2013wxa}
\be
p_\taupm^\mu \,=\, \frac{1 \mp a}{2} p_H^\mu  \,\pm\, \frac{b}{2} \ppip
\,\mp\, \frac{c}{2} \ppim \,\pm\, d q^\m \,.
\ee
The advantage of this expansion is that the constraint \eqref{eq:ptau_sum} is automatically satisfied.
We traded the remaining unknown 4-vector $(\ptaup - \ptaum)$ by
the four unknown coefficients, $a, b, c$ and $d$.
Our first goal is to determine these coefficients by solving 
four mass-shell constraints.

The first two mass-shell constraints are
$(p_\taup - p_\pip)^2 = m_\nu^2 = 0$,
and
$(p_\taum - p_\pim)^2 = m_\nu^2 = 0$.
They can be recast into 
\ba
m^2_\tau + m^2_\pi - x + a x - b m_\pi^2 + c z &=& 0 \,,
\nonumber \\
m^2_\tau + m^2_\pi - y - a y + b z - c m_\pi^2 &=& 0\,,
\label{eq:mass-shell-1}
\ea
where we introduced $[x, \, y, \, z] \equiv [ ( p_h \cdot p_\pip),\,(p_h \cdot p_\pim),\, (p_\pip \cdot p_\pim) ] $.
Similarly, the remaining two mass-shell conditions, 
$p_\taup^2 = m_\tau^2$ and $p_\taum^2 = m_\tau^2$,
can be written as
\ba
m_\tau^2 &=& \left( \frac{1 - a}{2} \right)^2 m_h^2 + \frac{b^2 + c^2}{4} m_\pi^2 + d^2 q^2 
\nonumber \\
&& + \frac{(1 - a)b}{2}  x
- \frac{(1 - a)c}{2}  y
- \frac{bc}{2} z \,,
\nonumber \\
m_\tau^2 &=& \left( \frac{1 + a}{2} \right)^2 m_h^2 + \frac{b^2 + c^2}{4} m_\pi^2 + d^2 q^2
\nonumber \\
&& - \frac{(1 + a)b}{2}  x
+ \frac{(1 + a)c}{2}  y
- \frac{bc}{2} z \,.
\label{eq:tau_mass_cond}
\ea
By subtracting these, we get
\be
a m_h^2 - bx + cy \,=\, 0 \,.
\label{eq:mass-shell-2}
\ee
The three equations in Eqs.~\eqref{eq:mass-shell-1} and \eqref{eq:mass-shell-2} can be organised in a matrix form 
\be
[{\bf M}] \cdot {\bf v} \,=\, {\bf \Lambda}\,,
\label{eq:mat_form}
\ee
with
\ba
[\bf M] \,=\,
\begin{pmatrix}
-x & m_\pi^2 & - z
\\
y & -z & m_\pi^2
\\
m_h^2 & -x & y
\end{pmatrix},
~
{\bf v} = 
\begin{pmatrix}
a \\ b \\ c
\end{pmatrix},
~
{\bf \Lambda}
\,=\,
\begin{pmatrix}
\lambda_x \\ \lambda_y \\ 0
\end{pmatrix},
\ea
and $(\lambda_x,\,\lambda_y) \,=\, (m_\tau^2 + m_\pi^2 - x,~m_\tau^2 + m_\pi^2 - y)$.
The solution can be readily obtained by inverting Eq.\ \eqref{eq:mat_form} as
\be
{\bf v} \,=\, [{\bf M}]^{-1} \cdot \bf \Lambda\,. 
\ee

The last coefficient, $d$, can be obtained by considering  
the sum of the two equations in \eqref{eq:tau_mass_cond}.
This leads to
\ba
d^2 &=& \frac{1}{- 4 q^2} \Big[ 
(1 + a^2) m_h^2 + (b^2 + c^2) m_\pi^2 - 4 m_\tau^2 
\nonumber \\
&& +\,2 (acy -abx - bcz) \Big]\,.
\ea
For physical solutions, the right-hand-side must be positive.
For positive $d^2$, there are twofold solutions for 
$p_\taup$ and $p_\taum$,
denoted by $p_\taup^i$ and $p_\taum^i$,
corresponding to $d > 0$ ($i=1$) and $d < 0$ ($i=2$), respectively.

\twocolumngrid
\bibliography{refs}

\end{document}